\documentstyle[12pt]{article}
\begin{document}
\hoffset 0.5cm
\voffset -0.4cm
\evensidemargin 0.0in
\oddsidemargin 0.0in
\topmargin -0.0in
\textwidth 6.7in
\textheight 8.7in

\begin{titlepage}

\begin{flushright}
PUPT-1689\\
hep-th/9703201\\
March 1997
\end{flushright}

\vskip 0.2truecm

\begin{center}
{\large {\bf A Note on the Transverse Five-Brane in M(atrix) Theory}}
\end{center}

\vskip 0.4cm

\begin{center}
{Gilad Lifschytz}
\vskip 0.2cm
{\it Department of Physics,
     Joseph Henry Laboratories,\\
     Princeton University, \\
     Princeton, NJ 08544, USA.\\ 
    e-mail: Gilad@puhep1.princeton.edu }

\end{center}

\vskip 0.8cm

\noindent {\bf Abstract}
 
We describe a way  to  compute scattering amplitudes
 in M(atrix) quantum mechanics,
that involve the transverse five-brane. We then compute certain
scattering processes and show that they have the
 expected $SO(5)$ invariance, 
give the correct transverse-five-brane mass, 
and agree with the supergravity result.

\noindent                
\vskip 6 cm

\end{titlepage}
\section{introduction}
Recently there has been a proposal that 
 M-theory \cite{wit1,ht} in the infinite momentum frame can be described
as the large $N$ limit of $SU(N)$ Yang-Mills quantum mechanics\cite{bfss}.
 In subsequent work
the brane \cite{dlp,gr,pol} 
content of type IIA theory have been discussed in this frame work
\cite{sus,grt,ab,gilmat,bss,gil3}\footnote{Branes in the IKKT \cite{ikkt}
model were discussed in \cite{li,cmz}}. The branes are represented as
 classical backgrounds.
For the transverse five-brane
or the type IIA $NS$-five-brane, one does not have such
 a representation.
In this note we will describe how to compute scattering
 amplitudes involving
transverse five-brane (we will often call it $5^{NS}$-brane), and by
that confirming its existance.

Starting with the $0+1$ theory (quantum mechanics),
 one can T-dualise to  other
$d+1$ dimensional theory and then use some specific duality
 of that field 
theory,  T-dual back and get some new information in the
 $0+1$ dimensional 
theory. Here we will employ this strategy in order to
 compute some properties
of the transverse five-brane. One does not have a description of the
transverse five-brane in $0+1$ dimensional theory, however as noticed
in \cite{grt} there is an indirect way of describing it. Start
with the configuration of the $(2+0)$-brane 
(sometimes called the membrane of matrix theory) in $0+1$-dimensions,
 and have extra three dimension
be compact (not necessarily  small) call them $X_1,X_2,X_3$.
 Now If we start
with the $(2+0)$ in the direction $(4,5)$ and T-dualise along
 $X_1,X_2,X_3$,
then we end up with a $(5+3)$-brane  in $3+1$ dimensions.
 Now one can use the 
S-duality\footnote{For a discussion of the validity of S-duality in the 
limit $N \rightarrow \infty$ see \cite{kss}}
 of $(3+1)$ Yang-Mills,
 which is equivalent to the S-duality of type IIB string to get to 
the configuration $(5_{NS}+3)$-brane,  then T-dual back to get the 
configuration $(5_{NS}+0)$-brane  in the $0+1$ dimensional theory
stretching in directions $(1-5)$. Unfortunately
we do not know how the S-duality acts on the commutator
 $[X_4,X_5]=ic$, so one
does not have the description of the transverse five-brane.

Let us instead focus on  some scattering amplitudes involving the 
transverse five-brane, as the procedure above takes us from $0+1$ theory
to another $0+1$ theory (albeit with possible different parameters)
 one can
 map a scattering amplitude involving the transverse five-brane to a known
amplitude, computable in $(0+1)$-dimensional matrix theory. 
In this fashion one can get certain 
information on the transverse five- brane in the context that we want.
 In order to do so 
all we need is to map known configurations in one theory to the other via
the transformation $T^3 S T^3$, and also to know how the parameters
 (coupling and
size of the space) change. Of course one needs to know
 the exact result in
the original theory in order to be able to do this as
 typically the coupling
will change, luckily we expect the long distance scattering
 to be exactly
computable in the matrix theory.

We start in section (2) with a review of T-duality in
 the context of matrix
theory, then in section (3) we describe how various configurations are
 transformed under the $T^3 S T^3$ transformation.
 We calculate in section (4)
scattering of zero-branes from the transverse five-brane
 ($5^{NS}+0$), and
show that it has an $SO(5)$ invariance and agrees
 with the super-gravity
result.From these calculation we extract the
 transverse-five-brane mass.
 We also give the results of the scattering of two Five-branes
and a five-brane and an anti-five-brane.

\section{Review of T Duality}
In this section we  review the description of compactification
 in matrix theory on 
a $d$ dimensional torus\footnote{Discussion of compactifications on other
manifolds can be found in  \cite{ks,dg,motl,kr,dos}}
.
We start with the $(0+1)$ dimensional field theory, If one of the 
nine direction is compact (with length $L$)
one has in the theory new sectors of states, namely
open strings that wind around the compact directions \cite{tay}.
If instead of describing those states this way one just Fourier transform
to another description, then one sees that the same theory is naturally
described as a $(1+1)$ dimensional theory with the extra dimension
being compactified on a circle of length  $= \frac{l_{s}^{2}}{L}$ 
($l_{s}=(\frac{l_{p}}{R_{11}})^{1/2}l_{p}$, 
where $R_{11}$ is the radius of the 
eleventh direction and $l_{p}$ is the Planck length in eleven dimensions.) 
Given that one has
$d$ dimensions compact (on a torus)
one ends with a $d+1$ dimensional super Yang-Mills
compactified on the dual d-torus \cite{tay}. 
Now this procedure is just the T-duality of type IIA (at strong coupling),
as replacing winding modes with momenta modes (Fourier transform) is
T-duality. In fact in the Lagrangian given a field $X_i$, if we 
Fourier transform (T-dualise)
in that direction then  \cite{grt}
\begin{equation}
X_i \rightarrow i\partial_{x_1} + A_{i} (x_i)=D_{i},
\end{equation}
where $x_i$ is now a parameter ranging on the dual circle and 
$A_{i}(x_i)$ captures the degrees of freedom in the $i$ direction
of the original brane (position, momentum).

Let us see how the T-duality acts on the branes of the theory. Starting
in $(0+1)$ dimensions the basic objects can be thought as
 zero-branes, while
in $(d+1)$ dimensions the basic objects are branes of dimension d, as
one expects. Starting with  a membrane configuration $[X_1, X_2]=ic$ in
$(0+1)$ dimension, if one T-dualise in directions other that
 $(1,2)$ then
we just end with a configuration $[X_1,X_2]=ic$ in a higher dimensional
super Yang-Mills theory. This configuration can easily be
 seen to describe
a ($d+2$)-brane (of course bound to many d-branes), as one
expects from T-duality.
If however we Fourier transform on coordinate $X_1$ then one ends in a 
$(1+1)$ dimensional filed theory with a configuration 
\begin{equation}
[i\partial_{x_1} + A_{i} (x_i), X_2]=ic \rightarrow 
D_{x_1} X_2 =ic.
\label{ss}
\end {equation}
Now as $x_1$ is the world sheet coordinate along the 1-brane
 (which is the
elementary object of this theory) this configuration describes a 1-brane
at an angle with respect to the $x_1$ direction. If we just T-dualise
the configuration of a membrane bounded to zero-brane in
 Type IIA string theory
then we will get a 1-brane with an angle, just like we have.
What we mean when we say a string at an angle is the following. As we are
on a compact space, a string configuration that is wrapped n times
in one direction and m times in another direction is a string at an angle
$\sim m/n$ for $n \gg m$, in the non compact space.

T-dualising now in the $X_2$ direction one ends up with
$[D_{1},D_{2}]=iF_{12}=ic$ a magnetic field on the world volume
 of a $2+1$
theory \cite{grt}. As is well known this represents zero-branes
bounded to the two-branes \cite{doug}, which is the T-dual of
 the configuration
we started with.

Let us now start with a configuration of a moving zero-brane in the $0+1$
dimensional theory. This is described by a background $X_1=vt$ which
can be recast in the form $[D_{t},X_1]=iv$, Fourier transform along
$X_1$ to get a configuration in $1+1$ dimensions of the form 
$[D_{t},D_{1}]=iF_{01}=iv$ which implies the existence of a 
constant electric field in the $1+1$ dimensional theory as one
 expects from
T-duality.  

Start with the configuration of a bound state of a four-brane bounded
to two-branes and zero-brane in the theory in $0+1$ dimensions. This is 
described by $[X_1,X_2]=ic_1,\ \ [X_4,X_3]=ic_2$ where
$c_1=\frac{2\pi R_1 R_2}{n_1}$ $c_2=\frac{2\pi R_3 R_4}{n_2}$, and the
number of zero-brane is $N= n_1 n_2$. The number of membrane
 in the $(1,2)$
direction is $\frac{1}{i2\pi R_1 R_2} Tr [X_1,X_2]=N c_1=n_2$ and the
number of membranes in the $(3,4)$ direction is $n_1$. Now
T-dualise along $X_1$ one gets a configuration in $1+1$, which is
$X_2=c_1 x_1, \ \ [X_3,X_4]=ic_2$. This describes in the matrix theory 
$n_1$ 3-branes bounded to N 1-strings in the $x_1$ direction and one
string at an angle  $\sim c_1$, exactly what one expects from the string 
theory.

\section{The Action of $T^3 S T^3$}
In this section we will describe how configuration change under the 
$T^3 S T^3$ transformation and how the parameters of the theory change.

\subsection{Mapping of Configurations}

In order to make things clear let us always call
 the three directions we T-dualise in
$X_1,X_2,X_3$, any different direction will be labeled as $X_i$.
As was explained before under the transformation $T^3 S T^3$ a $(0+1)$ 
dimensional
field theory is transformed back to a $(0+1)$ dimensional 
field theory.

Under this transformation a membrane, described by
 the classical background,
 $[X_i,X_j]=ic$ is transformed to a 
transverse five-brane in directions $(1,2,3,i,j)$.  Take a membrane
$[X_1,X_i]=ic$ under $T^3$ one gets $D_1 X_i =ic$
 ($c=\frac{2\pi R_i R_1}{N}$)
which is a configuration of  $N$ three-branes
in directions $(1,2,3)$ and one three-brane in directions $(2,3,i)$.
 Then acting
by $S$ one gets the same configuration and acting again by
 $T^3$ we are back
in a configuration of a membrane $[X_1,X_i]=ic$.
Similar transformation can be made on other configuration and we will
 just give the 
results:
\begin{itemize}
\item Membrane in $i,j$ $\leftrightarrow$ transverse five-brane in
 directions $(1,2,3,i,j)$,
or in symbols $2_{(i,j)}+0$ $\leftrightarrow$ $5^{NS}_{(1,2,3,i,j)}+0$
\item  $2_{(1,i)}+0$ $\leftrightarrow$ $2_{(1,i)}+0$
\item zero-brane moving in $i$ direction ($0 +v_i$)$\leftrightarrow$ same
\item  $2_{(2,3)}+0$  $\leftrightarrow$
zero-brane moving in direction $1$ ($0+v_{1}$)

\item $4_{(1,i,j,k)}+2_{(i,j)}+2_{(1,k)}+0$ $\leftrightarrow$ 
$5^{NS}_{(2,3,i,j,k)}+5^{NS}_{(1,2,3,i,j)}+2_{(1,k)}+0$
\item $4_{(1,2,i,j)}+2_{(1,i)}+2_{(2,j)}+0$ $\leftrightarrow$ same.
\item $4_{(1,2,i,j)}+2_{(1,2)}+2_{(i,j)}+0$ $\leftrightarrow$
 $4_{(1,2,i,j)}+v_{3}+
5^{NS}_{(1,2,3,i,j)}+0$
\item $4_{(1,2,3,i)}+2_{(12)}+2_{(3,i)}+0$ $\leftrightarrow$
 $1^{NS}_{i}+v_{3}+2_{(3,i)}+0$
\end{itemize}

Given the results for the potentials between these configurations given in
\cite{gilmat,gil3} and taking care to take into account the effect of
 the extra compact 
direction one can read off the potentials between configurations
 that one does not have
access directly in the $0+1$ dimensional field theory.

Now as $T$ and $S$ duality does not change the supersymmetry of a
 configuration we
can also see that we can get new configurations that preserve
 a quarter of the 
supersymmetry that has relative velocity between the brane
 configuration. For
example two stationary orthogonal $(2+0)$-branes is a
 supersymmetric configuration if
there parameter $c$ are the same. This implies that a zero-brane
 moving with a certain
velocity parallel to a $(5_{NS} +0)$ bound state is also supersymmetric.
 Similarly
because a configuration of $(4+2+2+0)$ and a stationary zero-brane is
 supersymmetric \cite{bdl,gil2}
for a specific choice of parameter, then all the configuration on the
 right-hand side
off the table above plus a zero-brane are
 supersymmetric\footnote{Some related
classical supergravity solutions can be found in \cite{tsy1,bmm,bmm1}}.

\subsection{ Parameter transformation}

Let us start with a string theory at coupling $g$, the  lengths for the
 $(1,2,3)$ directions
labeled by
$L_1,L_2,L_3$, and lengths for the other directions $X_j$. Under T-duality
in a $k$ direction the length $L^{'}_{k}= \frac{l_{s}^{2}}{L_k}$ and the
coupling $g^{'} = g\frac{l_s}{L_k}$. Under S duality $g^{'}=g^{-1}$ and
all lengths transform $L^{'}=L (g^{'})^{1/2}$. Starting with the above
 parameters under
$T^3 S T^3$ one ends up with
\begin{eqnarray}
g^{'} & = & g V_{p}^{1/2} \nonumber \\
L_{i}^{'} & = & L_{i} V_{p}^{-1/2} \nonumber \\
X_{i}^{'} & = & X_{i} V_{p}^{1/2} 
\label{var}
\end{eqnarray}
 Where $V_p=\frac{L_1 L_2 L_3}{g l_{s}^{3}}$ 
is the volume in Planck units of the space $L_1,L_2,L_3$.

Notice that we have used the S-duality transformation for the
 parameter from 
string theory, off course with the philosophy at hand this should come
 from the S-duality
symmetry of the Yang-Mills theory. In $(3+1)$  Yang-Mills under
 S-duality the 
three-torus stays the same, the coupling is inverted and the
 scalars are
 multiplied by 
$V_{p}$ (the scalars represent the transverse space).
 Using then the conformal invariance we scale the torus sides by
 $V_{p}^{1/2}$
which is accompanied multiplying the scalars by $V_{p}^{-1/2}$
 which is then
just the S-duality of the string theory. 

Due to the $T^3 S T^3$ symmetry of the theory the phase shift
 \cite{bac} of a
 configuration 
must be mapped to the phase shift of the mapped configuration.
\begin{equation}
A=-\int dt V(b^2+v^2 t^2)=A'=-\int dt' V'(b'^{2}+v^2 t'^{2})
\label{psh}
\end{equation}
where $b'^{2}=b^2 V_{p}$ and $t'=t V_{p}^{1/2}$.

\section{Transverse five-brane scattering}

In this section we will calculate various scattering process 
involving the transverse five-brane, and show that we get the correct
long distance result. 

We will calculate here the scattering of a zero-brane (or graviton)
 from the transverse five-brane. There are three different direction
 the zero-brane
can move. First its motion can be transverse to the five-brane,
 this is mapped
to a calculation of a zero-brane scattering off a membrane with
 velocity in the
$(6-9)$ directions. Second the zero-brane can move
 parallel to the five-brane.
This case is actually two-case that should agree.
 Movement along the ($1-3$) 
directions is mapped to a computation involving two
 relatively stationary
orthogonal two-branes, and movement along the $(4,5)$
 direction is mapped to a 
zero-brane moving parallel to a two-brane. The last two-computations 
although they are different should give the same answer due to the
 $SO(5)$ symmetry
of the transverse five-brane.

Now when computing the scattering one must take into
 account the three compact 
directions ($1,2,3$)\footnote{Compact brane in matrix
 theory were also considered
in \cite{bc}, and in string theory in \cite{gil2}}
.
\subsection{Zero-brane scattering off a transverse five-brane}

Starting with  a zero-brane scattering off a transverse five-brane
 with velocity transverse
to the five-brane, we calculate a zero-brane scattering off a
 two-brane (described by $[X_4, X_5]=ic$) with three 
orthogonal directions much smaller that the distance between the zero-brane 
and two-brane.  We get the long range potential \cite{gilmat}
\begin{equation}
V=-\frac{1}{2} \int \frac{ds}{s} e^{-b^2 s}
 \frac{(c^2 + v^2 )^2}{2c \sqrt{\pi}}s^4 
(\frac{\pi^{3/2}}{L_1 L_2 L_3 s^{3/2}})
\end{equation}
Where the last factor in parenthesis is from the compactified directions, 
 $c=\frac{L_4 L_5}{2\pi N}$, and $b$ is the distance between the
 membrane and the 
zero-brane.

Let us interpret this result using five-brane variables.
 and using equation (\ref{psh}).
 Labeling by prime, quantities after
the $T^3 S T^3$ transformation and using equation (\ref{var}) we find
\begin{equation}
c=\frac{L_4 L_5}{2\pi N}=\frac{L_{1}^{'}L_{2}^{'}L_{3}^{'}L_{4}^{'}L_{5}^{'}}
{2\pi N g^{'} l_{s}^{3}}=\frac{L_{1}^{'}L_{2}^{'}L_{3}^{'}L_{4}^{'}L_{5}^{'}}
{(2\pi)^{5/2} N g^{'} }
\label{c1}
\end{equation}
Further $ (2 \pi)^{3/2} g L_1L_2L_3=g'^2 (2\pi)^3$.

The potential between the zero-brane and transverse five-brane is then
\begin{equation}
V^{'}_{0,5}= -\frac{\pi (c^2+v^2)^2}{ 4 g' (2\pi)^{3/2} c}b'^{-2}
\label{v0t5}
\end{equation}

We will now calculate the scattering of a zero-brane off the
 transverse five-brane when
the zero-brane has velocity along direction $1$ (or directions $2$ or $3$).
the corresponding configuration is two orthogonal stationary $(2+0)$-branes.
The potential for this configuration was discussed in
 \cite{gilmat}. The potential
is
\begin{equation}
V=-\frac{1}{8\sqrt{\pi}} \int \frac{ds}{s^{3/2}}
\frac{e^{-b^2 s}}{\sinh c_1 s \sinh c_2 s}
[4+2\cosh 2c_1 s +2\cosh 2c_2 s -4\cosh (c_1-c_2)s -4\cosh (c_1+c_2)s ]
\label{2ort2}
\end{equation}
Where $c_1=\frac{L_2 L_3}{2\pi N_1}$ and $c_2=\frac{L_4 L_5}{2\pi N_2}$.
The long range potential when taking into account 
 one extra transverse (to both) compact
direction is,
\begin{equation}
V_{long}= -\int \frac{ds}{s^{3/2}}\frac{e^{-b^2 s}}{8\sqrt{\pi} c_1 s c_2 s}
(c_{1}^{2}-c_{2}^{2})^{2}
s^4 (\frac{\sqrt{\pi}}{L_1 s^{1/2}})=-\frac{(c_{1}^{2}-c_{2}^{2})^{2}}
{8 c_1 L_1 c_2 b^{2}}.
\end{equation}
Now after the $T^3 S T^3$ transformation the two-brane
 in the $(2,3)$ direction is
mapped to one unit of momentum along direction $1$,
 while the zero-branes
that were bound to it are mapped to zero-branes. The
 velocity of the cluster of $N_1$
zero brane will then have the velocity
\begin{equation}
v=\frac{2 \pi}{N_1 L_{1}^{'} T_{0}^{'}}
\end {equation}
Where $L_{1}^{'}$ is the length of direction $1$ after
 the transformation and
$T_{0}^{'}=\frac{2\pi}{g^{'} l_s}$ 
is the zero-brane mass after the transformation.
 Using equation (\ref{var})
one finds 
\begin{equation}
v=\frac{L_2 L_3}{2 \pi N_1}=c_1
\label{c}
\end{equation}
Thus the potential between a zero-brane moving along direction $1$ and a
 transverse five-brane stretched in directions $(1-5)$ 
\begin{equation}
V^{'}_{5,0} = -N_1\frac{\pi (v_{1}^{2}-c_{2}^{2})^{2}}
{4 (2\pi)^{3/2}c_2 g^{'}} b'^{-2}.
\label{50l1}
\end{equation}
Where $c_2=\frac{L_{1}^{'}L_{2}^{'}L_{3}^{'}L_{4}^{'}L_{5}^{'}}
{(2\pi)^{5/2} N g^{'} }$.

\subsection{}

If the zero-brane is moving in directions $(4,5)$ then this is
 mapped to a zero-brane 
moving in directions $(4,5)$ along a bound state
 of $(2+0)$-branes stretched
in directions $(4,5)$. So let us calculate the later configuration
 in the matrix theory.

In order to calculate the phase shift one has to
 calculate the one-loop vacuum
energy of the zero-brane quantum mechanics in the
 presence of a background
representing the membrane and the moving zero-brane.
We follow the notation of \cite{gilmat,gil3}.
 One needs to calculate determinants of the
operator $(-\partial_{0} +M^2)$, where $M^2$ is the mass squared 
of the off diagonal elements of the matrices.
The membrane is defined by the coordinates $X_5=P$ $X_4=Q$ and
$[Q,P]=ic=i\frac{L_4 L_5}{2\pi N}$,
 and the zero-brane is traveling with velocity $v$ in direction
$4$.
 Define 
\begin{equation}
H=P^2 + (Q-vt)^2 +Ib^2
\end{equation}
Then (in Euclidean space),
for the complex bosons one has six with $M^2=2H$ one with
$M^2=2H -4\sqrt{c^2-v^2}$ and one with $M^2=2H + 4\sqrt{c^2-v^2}$
For the fermions (in Euclidean space) one finds
\begin{equation}
M_{f}^{2}=H-ic \gamma_5 \gamma_4-iv \gamma_4
\end{equation}
thus giving eight fermions with $M_{f}^{2}= H-\sqrt{c^2-v^2}$ and eight 
with $M_{f}^{2}= H+\sqrt{c^2-v^2}$.

Evaluating the determinants
we find that the potential takes the form
\begin{equation}
V=-\frac{1}{4\sqrt{\pi}} \int \frac{ds}{s^{3/2}}
 \frac{e^{-b^2 s}}{\sinh cs}
[6+2\cosh 2\sqrt{c^2 -v^2} s -8\cosh \sqrt{c^2 -v^2} s ]
\end{equation}
This gives a long range potential (after taking into
 account three transverse compact
directions)
\begin{equation}
V_{long} =-\frac{\pi (c^2 - v^2)^2}{4 c L_1 L_2 L_3} b^{-2}
\end{equation}
We can now calculate the potential between a transverse
 five-brane and a cluster
of $N_1$ zero-branes moving with velocity in the $4$-th direction
\begin{equation}
V'_{5,0}=-\frac{N_{1} \pi (c^2 - v^2)^2}{4 c g' (2\pi)^{3/2}} b'^{-2}
\label{50l}
\end{equation}
We see that equations (\ref{50l1}) and (\ref{50l}) (remembering
$c=c_2$ ) agree thus confirming
the SO(5) invariance expected from the transverse five-brane .
We will shortly also compute the mass of the transverse five-brane.

\subsection{Supergravity calculation}

Let us compare this to a 
supergravity calculation of a scattering of a zero-brane off
 a transverse five brane
moving in the $11$ direction (i.e from the point of view of
 type IIA string theory, this
is the $(5_{NS}+0)$ bound state).
The metric for the transverse five brane in eleven-dimensions
 was given in \cite{rustsy}
(we slightly changed the notation),
\begin{equation}
ds^2=H^{1/3}[H^{-1}(-d\tilde{t}^2 +dy_{1}^{2}+dy_{2}^{2}+dy_{3}^{2}+
dy_{4}^{2}+dy_{5}^{2})
+d\tilde{y}_{11}^{2} +dx^{i}dx_{i}]
\end{equation}
Here $\tilde{t}=\frac{1}{\cos \theta}(t-\sin \theta y_{11})$,
 $\tilde{y}_{11}=\frac{1}{\cos \theta}(y_{11}-t \sin \theta )$,
$H=1+\cos^{2} \theta \frac{Q}{r^2}$, and $Q=\frac{M_{5}}{\cos \theta}$.

computing the potential for null geodesics \cite{gil1}
on this metric we find
\begin{equation}
V \sim -\frac{M_{5} (\tilde{P}^{2}_{11}+P^{2}_{\perp}) \cos^2 \theta}
{\cos \theta r^2} 
\label{sgv}
\end{equation}
Where
\begin{equation}
\tilde{P}_{11}=\frac{1}{\cos \theta}(P_{11}-E \sin \theta )
\label{p11}
\end{equation}
and $E,P_{11},P_{1}, P_{\perp}$
 are the energy, momentum in the $11$ direction,  momentum
in direction $1$, and momentum in the perpendicular direction
 of the scattered zero-brane, respectively.

We are interested in the case where $P_{\perp}=0$ but $P_1 \neq 0$.
 The velocity
in direction $1$ is then
\begin{equation}
\frac{E^2}{P_{11}^{2}}=1+\frac{v^2}{1-v^2}=1+(\gamma v)^2=\cosh^2 \nu
\end{equation}
The potential then becomes
\begin{equation}
V \sim -\frac{M_{5}(1-\sin \theta \cosh \nu)^2}{\cos \theta} r^{-2}
\label{potsug}
\end{equation}
To compare to the matrix calculation we should go to the limit
 where the membrane
velocity in the $11$ direction approaches $1$, and the velocity
 of the zero brane in 
direction $1$ is small. In this case we write $\theta= \pi/2 -c'$.
The potential at this limit is
\begin{equation}
V \sim -\frac{M_{5}((c')^2 - v^2)^2}{c'} r^{-2}
\end{equation}
This is the same as equations (\ref{50l1}) and (\ref{50l}) if we identify
$c=c'$.

Let us determine $c'$. From the supergravity solution one has
\begin{eqnarray}
\sin \theta & = & \frac{P_{11}}{E}=\frac{NM_0}{E} \\
E^2 & = & M_{5}^{2}+ N^2 M_{0}^{2}
\end{eqnarray}
Where $E, P_{11}$ are the energy, and  the  momentum in the $11$ 
direction of the moving five-brane (which is the eleven dimensional
description of the bound state of a transverse five-brane and zero-branes)
and $M_{5},M_{0},N$ are the mass of the five-brane mass of a zero-brane
 and the number
of zero-brane respectively.
Then we find
\begin{equation}
c'=\frac{M_{5}}{N M_{0}}
\end{equation}
From this identification and the identification from 
the scattering in string theory 
$c=c'$  we can read off the 
transverse five-brane mass in the matrix calculation ($l_{s}^{2}= 2\pi$).
\begin{equation}
M_{5} =\frac{L_{1}^{'}L_{2}^{'}L_{3}^{'}L_{4}^{'}L_{5}^{'}}
{(2\pi)^2 (g^{'})^2}
\end{equation}
Which is the correct transverse five-brane mass.

If we now calculate the scattering in the  supergravity with
 $P_{\perp} \neq 0$
and $P_{1}=0$ we will get the result from the matrix model equation
 (\ref{v0t5}).
The agreement may also be viewed as an additional evidence for S-duality.

Notice that if we wanted to check that an elementary string will acquire 
a Berry phase once it is transported around the $NS$ five-brane,
this would correspond to the transportation of a four-brane
 around a two-brane
in the original theory which does acquire the berry-phase \cite{bd}.

\subsection{}

Similar calculations can be done for the 
 scattering of two transverse five-branes ($5^{NS}+0$),
 this is mapped to a computation of scattering two membranes ($2+0$).
Taking care of the compact directions,
the resulting potential between two membranes and between
 a membrane and an 
anti-membrane are \cite{gilmat}
\begin{eqnarray}
V_{2,2} & = & -\frac{L_4 L_5 v^4}{8c^2 L_1 L_2 L_3 }r^{-2} \\
V_{2,\bar{2}} & = & -\frac{ 2 L_4 L_5 c^2}{L_1 L_2 L_3} r^{-2}
\label{mm}
\end{eqnarray}
Using  the transformation $T^3 S T^3$ this gives the potential between 
two moving transverse five-branes in matrix theory,
 and the potential between 
a transverse five-brane  and an anti transverse five-brane,
 to be ($l_{s}^{2} =2\pi$) 
\begin{eqnarray}
V'_{5,5} & = & -\frac{L^{'}_{1} L^{'}_{2}L^{'}_{3}L^{'}_{4}L^{'}_{5}
}{g'^2 (2 \pi)^{5/2}}
\frac{v^4}{4c^2 \pi^{1/2}}r'^{-2} \\
V'_{5,\bar{5}} & = & -\frac{L^{'}_{1} L^{'}_{2}L^{'}_{3}
L^{'}_{4}L^{'}_{5}}
{g'^2 (2 \pi)^{5/2}}
\frac{4 c^2}{\pi^{1/2}}r'^{-2}
\end{eqnarray}
and $c$ is as in equation (\ref{c1}).
Off course one can continue and calculate scattering processes
 with all the states 
described in section (3), and probably others.

In general 
this way of calculating scattering amplitudes is no less powerful
 than the way one does
that for configurations which we know how to represent
 them as classical backgrounds. However observing the connection between
parameters given in section (3.2), one sees that we are unable to 
describe the limit in which the five-brane are un-compactified. 
This maybe related to the absence of the corresponding charge in the
supersymmetry algebra of zero-branes in \cite{bss}.

\begin{center}
{\bf Acknowledgments}
\end{center}
I would like to thank S.D. Mathur and S. Ramgoolam for helpful discussions.

\end{document}